\documentclass{article}
\usepackage{graphicx}
\usepackage{amsmath, amssymb, graphics, setspace}
\usepackage{amsmath}   
\usepackage{bm} 
\usepackage{dcolumn}
\usepackage{color}
\usepackage{xcolor}
\usepackage{mathrsfs}
\usepackage{xfrac}
\usepackage{amsfonts}
\usepackage{varioref}
      \usepackage[skip=0.333\baselineskip]{caption}
\usepackage{subcaption}
        \usepackage{lmodern}
        \usepackage{caption}
         \usepackage{floatrow}%
         \usepackage[english]{babel}
\usepackage[utf8]{inputenc}
\usepackage{mathtools}
\usepackage{placeins}
\usepackage[T1]{fontenc}
\usepackage{lipsum}
\usepackage{csquotes}
\usepackage{siunitx}
\usepackage[export]{adjustbox} 
\usepackage{bm}
\usepackage{ragged2e}
\usepackage{hyperref}
\hypersetup{colorlinks=true,citecolor=blue, linkcolor=blue, filecolor=blue, urlcolor=blue,}

\input epsf

\labelformat{section}{Section #1} 
\labelformat{subsection}{Section #1} 
\labelformat{subsubsection}{Section #1}
\labelformat{subsubsubsection}{Section #1}
\labelformat{equation}{Eq.~(#1)} 
\labelformat{figure}{Fig.~#1} 
\labelformat{subfigure}{Fig.~\thefigure#1} 
\labelformat{table}{Tab.~#1} 
\labelformat{appendix}{Appendix #1}
\usepackage{amssymb}
\title{\textbf{Novel modulus stabilization mechanism in higher dimensional   \textit{f}(\textit{R}) gravity}}
\author{Shafaq Gulzar Elahi
\footnote{ug18sge@iacs.res.in},
Soumya Samrat Mandal
\footnote{ug18ssm@iacs.res.in}\\
Soumitra SenGupta
\footnote{tpssg@iacs.res.in}\\\\
{\small{School of Physical Sciences, Indian Association for the Cultivation of Science, Kolkata-700032, India}}}

\begin{document}
\maketitle
\begin{abstract}
Stabilization of the moduli in any higher dimensional model is essential to obtain the lower dimensional effective theories where the stable values
of the moduli appear as parameters that in turn determine various observables in the present-day universe.
In this work, we obtain a new warped solution for a 5-dimensional $\text{f(R)} $ higher curvature gravity in an anti de-Sitter bulk. The higher curvature 
term appears as a natural generalization in the bulk gravity action and is shown to modify the usual warped metric. The novel feature of this modification leads to a geometric stabilization of the modulus/radion field in the underlying effective theory on the visible 3-brane without the need for any external stabilizing field. It is further shown that the stabilized value of the modulus resolves the well-known gauge hierarchy problem without any unnatural fine-tuning of the model parameters. This new solution along with the stabilized modulus  opens up the possibilities of new 
observable signatures in the effective lower dimensional theory where the stabilized value of the modulus  not only appears as a parameter in the lower dimensional brane  but also decides the dynamics of the modulus i.e. the radion.
\end{abstract}
\section{Introduction}\label{Intro}
Search for extra spatial dimensions has been a subject of active research  interest for a long time. 
Ever since Kaluza and Klein proposed a possible interpretation of electromagnetism  through extra spatial dimension, there has been  a plethora of work  exploring this feature extensively in different contexts. This ranges from issues related to small scale Physics such as the resolution of  gauge hierarchy/fine tuning  problem\cite{fine tuning}, the origin of neutrino masses\cite{neutrino},\cite{neutrino2}, fermion mass hierarchy\cite{fermion} and dark matter  to large scale phenomena such as inflation\cite{inflation},\cite{inflation2}, bouncing\cite{bounce}-\cite{bounce3} phenomena in cosmology as well  galactic structure\cite{galaxy}-\cite{galaxy3} in Astrophysics.  Moreover, the predictions of the inevitable existence of extra dimension in the context of string theory  generated intense activities to unearth  such an exotic but hidden feature of space-time geometry. Two extra dimensional  models namely large extra dimensions\cite{large}-\cite{Arkani-Hamed:1998jmv} and warped extra dimensions\cite{warp}-\cite{warp3}became extremely popular in the beginning of this century. The testing beds of these models ranged from collider Physics to cosmological/Astrophysical scenarios. 
In all these models one of the key signatures of the extra dimensions on a lower dimensional  hypersurface (branes) has been the moduli fields originating from various  metric components in higher dimensions.  In the context of the fine-tuning problem related to the  large radiative corrections to Higgs mass, the warped geometry model a la Randall and Sundrum\cite{Randall:1999ee}  was particularly turned out to be very  successful as it could resolve the problem without introducing any intermediate scale in theory.
Interestingly String theory indeed  can provide an analog of such a warped extra-dimensional scenario through a  throat-like  geometry\cite{KS} and 
Randall Sundrum model can capture the essential features of this throat geometry in a simple way so that possible  signatures of extra dimensions  in collider Physics can be estimated through various graviton KK modes\cite{KK}-\cite{KK3} with a much larger coupling ( Tev$^{-1}$couplings) with the standard model fields.
However, the stabilization of these moduli fields to their respective minima has been  a key feature in order to extract an acceptable Physics on our Universe (3-branes). In the low energy effective theory in lower dimensional brane various parameters depend crucially on the stabilized values of the moduli. Moreover, the fluctuation around this stable value of the modulus leads to the dynamical radion field whose interactions with the brane fields play a key role in determining the signature of extra dimension in collider Physics. As a result the mechanism of stabilizing the hidden world of extra dimensions always occupies the center stage of any theories with extra dimensions.  In this context, irrespective of the models, one requires additional fields to generate an appropriate potential term for the moduli so that the moduli can be stabilized to a desired value. The origin of such a stabilizing field is often unknown. In particular, for the warped geometry model, it was shown by Goldberger and Wise\cite{GW,GW2} that a bulk scalar field can ensure the modulus  stabilization for appropriate choices of the scalar parameters. Such stabilization was further generalized by Csaki et,al\cite{Csaki} by taking the back reaction of the scalar on the bulk geometry into consideration. However as discussed, the origin of such stabilizing field remained unexplained.  In the context of String theory also it was shown that flux of an external tensor field \cite{flux} can result in moduli stabilization. However, the string landscape leads to the well-known swampland conjecture where a large class of vacuua in the effective theories turns out to be inconsistent. In particular Ads bulk in general is accompanied by inherent instability. In the present discussion, we work with an Ads bulk just as the RS model which was primarily
invoked as a model to address the gauge hierarchy problem and showed great promise to yield interesting phenomenology and cosmology at the Tev scale provided the only modulus can be stabilized to a desired value without introducing any intermediate scale or fine-tuning. Thus following the original work of Randall and Sundrum, here also we focus our attention on a five-dimensional model with an Ads bulk in the presence of higher curvature terms. We will further comment on the swampland conjecture in the context of our work in a forthcoming section.
In this work, we propose a novel  mechanism to stabilize the modulus in a 5-dimensional warped geometry model without the need for any externally employed field. Here we exhibit that  the higher curvature geometry in the  bulk results in modulus stabilization without the need of invoking any external bulk scalar field by hand. In the context of higher curvature $\text{f(R)}$
model\cite{nojiri}-\cite{f3}, it has been demonstrated earlier that the scalar field in the dual scalar-tensor model of the original bulk $\text{f(R)}$ in principle can  lead to such stabilization of modulus \cite{ssg1}-\cite{ssg4}. However, that only implies the stabilization of a conformally transformed metric and does not correspond to the original warped metric one uses to estimate the signatures of the moduli on the lower dimensional hypersurface. Our work here, on the contrary, generalizes the original RS model with higher curvature $\text{f(R)}$ model. The inclusion of this bulk higher curvature term  is arguably more realistic  as the bulk spacetime in this model is endowed with a  negative cosmological constant of the order of the Planck scale.  We derive a new warped metric in the form of  perturbative corrections to the original RS solution where the correction is generated due to the higher curvature terms. We then explicitly show that such a geometry not only explains the gauge hierarchy problem but also leads to natural modulus stabilization without the need for any external stabilizing field. The additional degree of freedom associated with the higher curvature term in the background geometry is responsible for this  natural  stabilization of the modulus without the need for any external field. We once again emphasize that without modulus stabilization, no higher dimensional theory will have any significance in the lower dimensional universe since the ground state of the modulus and fluctuations around that determine all the key signatures and significance of a higher dimensional model. There lies the importance of our findings in the context of modulus stabilization in a natural geometric way.
We describe our higher curvature 5-dimensional model in  section 2 and then in section 3 we obtain a new warped geometric solution and also the brane tensions  using perturbative method. Section 4  elaborately  describes  the implications of our results along with a derivation of the radion action and  the corresponding potential.  In section 5, we conclude with a discussion of our result and on various windows of  future work in both small and large-scale Physics that are opened from our new warped solution with  a naturally and geometrically stabilized modulus.

\section{Framework: Braneworld Scenario in \text{f(R)} Theory}\label{framework}

The Randall Sundrum model considers a non-factorizable five-dimensional metric, with the 4D flat metric multiplied by an exponential warp factor which is a function of the extra dimension. The nature of the extra dimension $\phi \in [-\pi,\pi]$ is angular and is subjected to $S^1/Z_2$ orbifolding with fixed points 0 and $\pi$ identified.There are two 4-D flat branes located at the orbifold fixed points $\phi$=0 (Planck Brane)and $\phi$=$\pi$ (Visible/TeV Brane) and the bulk has a cosmological constant $\Lambda$.
The five-dimensional action is:
\begin{equation}
\mathcal{S}=\int{ d^5x\,\sqrt{-g}(\text{M}^3\text{R}-\Lambda)}-\int {d^4x \,\sqrt{-g_i}\mathcal{V}_i}
\end{equation}
where $\text{M}$ is the fundamental 5-D mass scale (we are working in natural units), $\text{R}$ is the 5-D Ricci Scalar,$\Lambda$ is the bulk cosmological constant, $\cal{V}$$_i$ is the tension of the ith brane(i=hid(vis)) and $\eta_{\mu\nu}$is the 4D metric.\\
The RS solution implies a negative bulk cosmological constant(the bulk is $AdS_5$), $\Lambda=-24\text{M}^3k^2$ and brane tensions $\cal{V}$${}_{\text{hid}}=-$$\cal{V}$${}_{\text{vis}}=24\text{M}^3k^2$.The Solution is:\\
 \begin{equation} \label{(2)} ds^2=e^{-2k r_c|\phi|}\eta_{\mu\nu}dx^{\mu}dx^{\nu}+r_c^2d\phi^2\end{equation}

Throughout the paper, we adopt the mostly positive metric convention. The size of the extra dimension is set by the compactification radius $r_c$ which is arbitrary here(not set by any dynamics) and unless one provides a mechanism to dynamically generate the compactification radius $r_c$ that resolves the Gauge Hierarchy Problem, the R-S Solution is considered incomplete. This is particularly alarming when one writes the Effective Field Theory on the brane.
The modulus field must therefore be stabilised.
Here we explore the stabilisation from the perspective of $\text{f(R)}$ gravity in the bulk. In the original RS model, the value for the bulk cosmological constant was chosen to be of the order of the Planck scale. This motivates us to work with higher curvature terms in the
the bulk which has significant contributions only on a high scale.
For a generalised $\text{f(R)}$ theory in D=5 brane-world scenario ,the action is given by:
\begin{equation}
\label{(3)}
\mathcal{S}= \text{M}^3 \int{ d^5x\,\sqrt{-{g}}~\text{f(R)}}-\int {d^4x \,\sqrt{-g_i}\mathcal{V}_i}
\end{equation}
$\Lambda$ is included in $\text{f(R)}$. From now on,we shall set $\text{M}$=1 for ease of calculations.\\

We are interested to study the nature of warped solutions when the bulk is described by $\text{f(R)}$.
Consider the RS like metric ansatz
\begin{equation}
\label{(4)}
ds^2=e^{-2\text{A(y)}}\eta_{\mu\nu}dx^\mu dx^\nu +\text{B(y)}^2dy^2
\end{equation}
where, y=$r_c \phi$

The field equations describing a general $\text{f(R)}$ theory are:
\begin{equation}
\label{(5)}
\text{R}_{\text{MN}} \text{f}_\text{R}(\text{R})-\frac{1}{2} g_{\text{MN}} \text{f(R)}+g_{\text{MN}} \square  \text{f}_\text{R}(\text{R})-\nabla_{\text{M}} \nabla_{\text{N}} \text{f}_\text{R}(\text{R})=\frac{1}{2}\textbf{T}_{\text{MN}}
\end{equation}

The indices $M$,$N$ run from (0,1,2,3,5) where 5 denotes the extra-dimension.$\text{f}_\text{R}(\text{R})$ denotes derivative of $\text{f(R)}$ with respect $\text{R}$, and the energy-momentum tensor is given by
$$\textbf{T}_{\text{MN}}^{(m)}=\frac{-2}{\sqrt{-g}} \frac{\delta \mathcal{L}_{m}}{\delta g^{\text{MN}}}$$
where  $\mathcal{L}_{\mathrm{m}}$ stands for the Lagrangian corresponding to the matter of some kind.\\
For this metric, the non-zero Christoffel symbols are: $\Gamma^5_{55}=\text{B(y)}'/\text{B(y)}$,$\Gamma^5_{\mu\nu}=\eta_{\mu\nu}e^{-2\text{A(y)}}\text{A(y)}'/\text{B(y)}^2$ and $\Gamma^\mu_{5\nu}=-\delta^\mu_\nu \text{A(y)}'$.
\begin{align}
   & \text{R}=\frac{-4}{\text{B}(y)^3}\left[2 \text{A(y)}'\text{B}'(y)+\text{B}(y)\left(5 \text{A(y)}'^2-2 \text{A(y)}''\right)\right]\label{(6)}\\
    &\text{R}_{55}=-4\left[\frac{\text{A}'(y)\text{B}'(y)}{\text{B}(y)}-\text{A}''(y)+\text{A}'(y)^2\right]\label{(7)}\\
    &\text{R}_{\mu\nu}=\frac{1}{\text{B}(y)^3}\eta_{\mu\nu}e^{-2\text{A(y)}}\left[\text{B}(y)\left(\text{A}''(y)-4 \text{A}'(y)^2\right)-\text{A}'(y)\text{B}'(y)\right]\label{(8)}
\end{align}
Considering a  leading order higher curvature correction,  we propose to solve  \ref{(5)}  for $\text{f(R)}=\text{R}+\alpha \text{R}^2-\Lambda$ ($\alpha$ is dimensionless).
Using the metric ansatz \ref{(4)} in \ref{(5)},we obtain the $\text{f(R)}$ gravity equations of motion (for the bulk): 

\begin{equation}
\label{(9)}
\begin{split}
&6 \text{A}'(y)^2-6 k^2 \text{B}(y)^2+\alpha  \left[-\frac{160 \text{A}'(y)^2 \text{B}'(y)^2}{\text{B}(y)^4}+\frac{64 \text{A}'(y)}{\text{B}(y)^3} \left(2 \text{A}''(y) \text{B}'(y) +\text{A}'(y) \text{B}''(y)-4 \text{A}'(y)^2 \text{B}'(y)\right)\right.\\
&\left.-\frac{8}{\text{B}(y)^2} \left(-4 \text{A}''(y)^2+5 \text{A}'(y)^4+8 \text{A}'''(y) \text{A}'(y) -32 \text{A}'(y)^2 \text{A}''(y)\right)\right]=0
 \end{split}
\end{equation}

\begin{equation}
\label{(10)}
\begin{split}
&\frac{1}{\text{B}(y)^7}\Biggl\{3 \text{B}(y)^4 \text{A}'(y) \text{B}'(y)+\text{B}(y)^5 \left(6 \text{A}'(y)^2-3 \text{A}''(y)\right)-6 k^2 \text{B}(y)^7+\textbf{$\alpha$} \Bigl[  -240 \text{A}'(y) \text{B}'(y)^3\\
&-80 \text{B}(y) \text{B}'(y) \left(3 \text{B}'(y) \left(2 \text{A}'(y)^2-\text{A}''(y)\right)-2 \text{A}'(y) \text{B}''(y)\right)-8 \text{B}(y)^2 \left(12 \text{A}'''(y) \text{B}'(y)+8 \text{A}''(y) \text{B}''(y)\right.\\
&\left.-16 \text{A}'(y)^2 \text{B}''(y)+37 \text{A}'(y)^3 \text{B}'(y)+2 \text{A}'(y) \left(\text{B}'''(y)-36 \text{A}''(y) \text{B}'(y)\right)\right)\\
&-8 \text{B}(y)^3\left(-2 \text{A}''''(y)+12 \text{A}''(y)^2+5 \text{A}'(y)^4+16 \text{A}'''(y) \text{A}'(y)-37 \text{A}'(y)^2 \text{A}''(y)\right)\Bigr]\Biggr\}=0\\
\end{split}
\end{equation}
\section{Perturbative Approach}\label{PA}
    Consider $\alpha$ to be small such that $ \text{R}^2$ in $\text{f(R)}$ is a small correction over $\text{R}$ solution, In this backdrop, consider the following ansatz for $\text{A(y)}$ and $\text{B(y)}$:
\begin{align}
     \text{A(y)}=& k~y+\alpha  \text{A}_1(y) \label{(11)}\\
    \text{B(y)}=& 1+\alpha\text{B}_1(y)\label{(12)}
\end{align}
 
Here, $k~y$ denotes the unperturbed RS-like part of the warp factor, with $k=\sqrt{-\Lambda/12}$  and $\text{A}_{1}$ and $\text{B}_{1}$ are the first order perturbative corrections respectively. Substituting the above ansatz for $\text{A(y)}$ and $\text{B(y)}$ into \ref{(9)} and \ref{(10)}, and considering only upto the leading order correction in $\alpha$ we get:\\

\begin{align}
   & -40 k^4 - 12 k^2\text{B}_1(y) + 12 k \text{A}_1'(y)=0 \label{(13)}\\
&  -3 \text{A}_1''(y)+12 k \text{A}_1'(y)+3 k \text{B}_1'(y)-12 k^2 \text{B}_1(y)-40 k^4=0\label{(14)}
\end{align}

Solving \ref{(13)} and \ref{(14)}, 
\begin{align}
    &\text{A}_1(y)= \frac{1}{2} b_0 k y^2+ \frac{10k^3}{3}y\label{(15)}\\
    &\text{B}_1(y)=b_0 y\label{(16)}
\end{align}

In principle, one can explore higher-order corrections as well, however, since we have considered $\alpha$ to be small we restrict ourselves to this order.\\
Therefore,
\begin{align}
&\text{A(y)}=\Tilde{k} y + \beta y^2 \label{(17)} \\
&\text{B(y)}=1+ \alpha b_0 y \label{(18)} 
\end{align}
where $\Tilde{k}=k +\frac{1}{3}10\alpha k^3$ and $\beta= \frac{1}{2}\alpha b_0 k$
\subsection{Brane Tensions}
We have solved \ref{(10)} for the bulk and obtained \ref{(17)} and \ref{(18)}. Now, we impose the $Z_2$ symmetry at the boundaries where the branes are located. We want to calculate the leading order corrections to the warp factor in this setup. Consider $\cal{V}$$_{hid(vis)}=$$\cal{V}$${}^0_{hid(vis)}+\alpha$$ \cal{V}$${}^1_{hid(vis)}$, where $\cal{V}$${}^0_{hid(vis)}$ are the tensions of hid(vis) brane in RS model.
Solving \ref{(10)} in the presence of branes(upto leading order), we obtain:
\begin{equation}
\label{(19)}
  \mathcal{V}_{\text{hid}}=-\mathcal{V}_{\text{vis}}= 12 k+~40\alpha k^3
\end{equation}

The brane tensions become more positive(negative) respectively.
\section{4-D Effective Theory}
Consider the low energy fluctuations about the solution \ref{(4)}
\begin{equation}
\label{metric-fluc}
ds^2=e^{-2~A(|\phi |, x)}g_{\mu\nu}(x)dx^\mu dx^\nu +\text{B}(|\phi|,x)^2 T(x)^2 d\phi^2
\end{equation}
The 5-D scalar curvature for this metric is
\begin{align}\label{scalar-curvature}
\rm R= &\rm (1+\alpha  b_0T(x)|\phi|)e^{2T(x)|\phi| (\Tilde{k} +\beta T(x)|\phi|)}~{}^4\mathcal{R} -(\partial_\mu T(x))^2\frac{2|\phi|~ e^{2 ~T(x)~ |\phi|(\Tilde{k} +\beta  ~T(x)~ |\phi|)}}{T(x)~ (1+\alpha  b_0T(x)~ |\phi|)^3} \mathcal{K}(x,|\phi|) \\ \nonumber
&-\rm \partial_\mu\partial^\mu T(x)\frac{2 e^{2 T(x) |\phi| (\Tilde{k} +\beta  T(x) |\phi|)}}{T(x) (1+\alpha  b_0 T(x) |\phi|)}\mathcal{P}(x,|\phi|)
-\mathcal{Q}(x,|\phi|)
\end{align}
where the functions $\cal{K}$$(x,|\phi|)$$,\cal{P}$$(x,|\phi|)$and $\cal{Q}$$(x,|\phi|)$ are defined in \ref{5D-Ricci-Definitions}
Here we no longer have a constant curvature solution, unlike RS. As $\alpha\to 0$, we recover the RS limit. We will use \ref{scalar-curvature}and \ref{metric-fluc} to derive the effective action for the modulus field, which will contain a non-minimally coupling between $T(x)$and the 4-D scalar curvature, kinetic terms, and the potential energy for the modulus field. We will study the scale of gravitational interaction and whether we can resolve the gauge hierarchy issue in the backdrop of this model. We will show that both these can be achieved without any hierarchy in the fundamental parameters of the theory. We will show that the higher curvature degrees of freedom will generate a potential for the modulus field with a minimum at such $r_c= <T(x)>$ which ensures the resolution of gauge hierarchy and that the fundamental scale of gravity remains unaffected. 

\subsection{Strength of Gravitational Interaction}\label{4Dgravity}
\begin{figure}[ht]
  {\includegraphics[height=1.8 in]{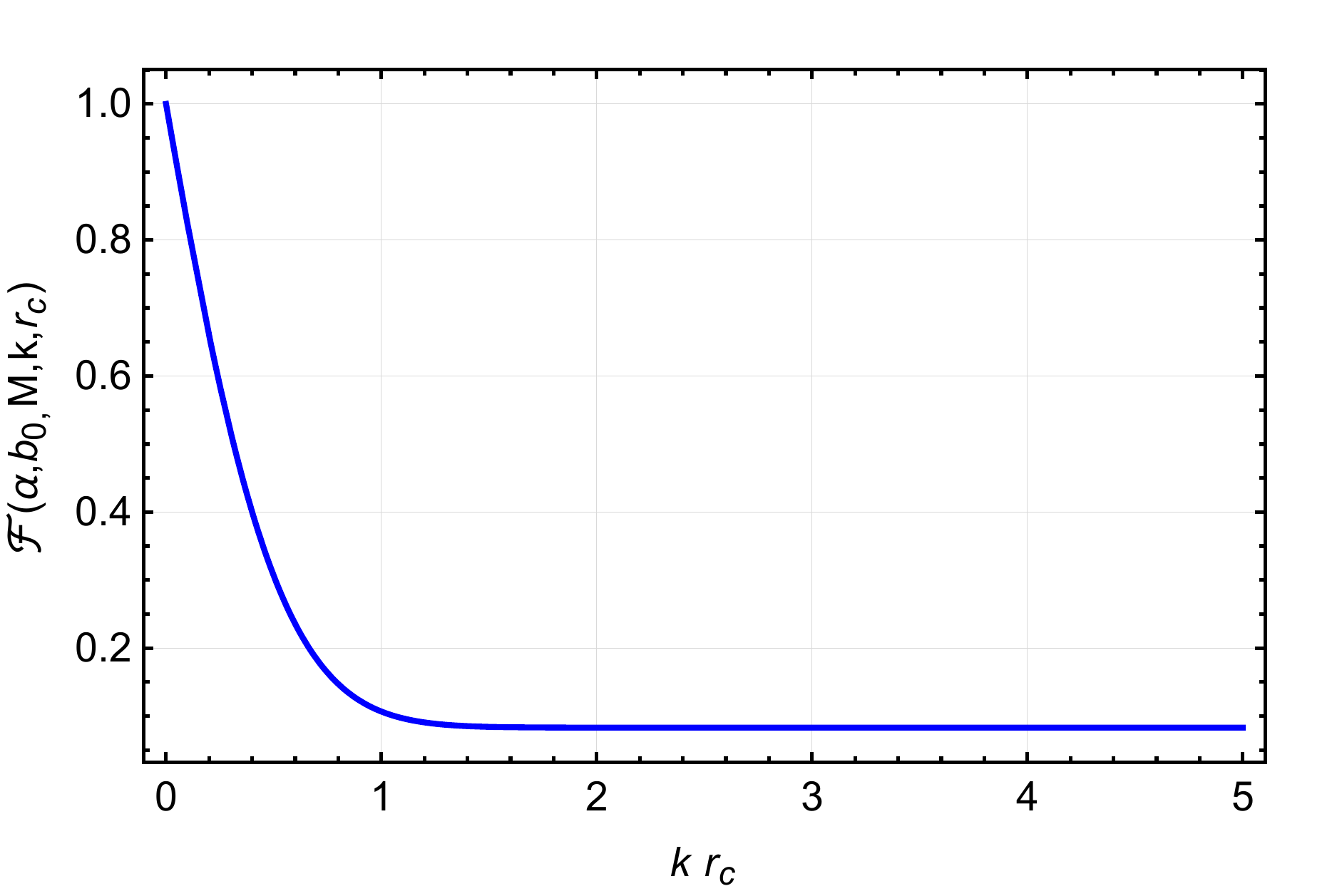}}
   \caption{Parameters:$~\alpha=0.489,\Lambda=-1$ and $b_0=1.19$}
   \label{5DPlanck}
\end{figure}

In order to determine the strength of gravitational interaction on the visible brane, we consider the effective action(reintroducing $\text{M}$)(assuming the modulus is stabilized):
\begin{align}\label{21}
 \mathcal{S}_{\text{eff}} \supset & \int d^4 x \int^{\pi}_{-\pi } r_c d \phi ~ \text{M}^3 \sqrt{-\overline{g}} ~e^{2A(|\phi | r_c)} {}^4\mathcal{R}\\
= & \int d^4 x \int^{\pi}_{-\pi } r_c d \phi ~\text{M}^3 \text{B}(|\phi|r_c) e^{-2 A(|\phi|r_c)}  \sqrt{-g} ~{}^4\mathcal{R}
 \end{align}
here $\rm r_c= <T(x)>$, $^{(4)}$$\cal{R}$ denotes the four-dimensional Ricci scalar made out of ${g}_{\mu \nu}(x)$, in contrast to the five-dimensional Ricci scalar, $\text{R}$, made out of $ \overline{g}_{\mu\nu} =e^{-2A(|\phi| r_c)}g_{\mu\nu}(x)$. 
Therefore, we can see that the 4-D and 5-D Planck Scales are related:
\begin{equation}\label{scales}
    \text{M}_{Pl}^2=\frac{\text{M}^3}{k}\left[1-\mathcal{F}(\alpha,k,r_c,\text{M},b_0)\right]
\end{equation}
where
\begin{equation}\label{22}
   \mathcal{F}(\alpha,k,r_c,\text{M},b_0)=e^{\Tilde{P}(r_c,\alpha,k,\text{M},b_0)}-\frac{1}{3 \sqrt{b_0}}10 \sqrt{\pi } \sqrt{\alpha } k^{5/2} e^{\Tilde{Q}(r_c,\alpha,k,\text{M},b_0)}\Big\{\textbf{erf}\left[\Tilde{N}(r_c,\alpha,k,\text{M},b_0)\right]-\textbf{erf}\left[\Tilde{S}(r_c,\alpha,k,\text{M},b_0)\right]\Big\}
\end{equation}
where $\textbf{erf}(x)=\frac{2}{\sqrt{\pi}}\int^{x}_0 dt e^{-t^2}$ and the expressions for $\Tilde{P},\Tilde{Q},\Tilde{N}$ and $\Tilde{S}$ can be found in \ref{4D5DAppendix}.From \ref{5DPlanck}, we can see for $k~r_c>1$,$\cal{F}$$(\alpha,k,\text{M},b_0,r_c)\to$~zero and hence 
$\text{M}_{Pl}\approx \text{M}$ similar to the case of \cite{Randall:1999ee}.
\FloatBarrier
\vspace{1mm}

\subsection{Physical Mass Scale}
The matter fields on the visible brane couple to the low energy gravitational field  $\overline{g}_{vis_{\mu \nu}}(x)=e^{-2 A (\pi r_c)}g_{\mu\nu}(x)$. The Lagrangian for the Higgs Field can be written as (set $\rm M=1$ again):
\begin{equation}\label{HiggsAction}
\mathcal{S}_{\text{vis}} \supset \int d^4 x \sqrt{-\overline{g}_{vis}} \Bigl\{ \overline{g}_{vis}^{\mu \nu} D_{\mu}
 \mathcal{H}^{\dagger} D_{\nu} \mathcal{H} - \lambda (|\mathcal{H}|^2 - v_0^2)^2 \Bigr\}
\end{equation}
The canonical Higgs field will be written as $e^{ A( r_c \pi)}$$ \cal{H} $
\begin{equation}\label{HiggsAction2}
    \mathcal{S}_{eff} \supset \int d^4 x \sqrt{- g}  
\Bigl\{ g^{\mu \nu} D_{\mu}
 \mathcal{H}^{\dagger} D_{\nu} \mathcal{H} - \lambda \left(|\mathcal{H}|^2 - e^{-2 A( r_c \pi)} v_0^2\right)^2 \Bigr\}
\end{equation}
Therefore, the physical mass scale on the visible brane is given by:
\begin{equation}
\label{hierarchy}
\textbf{v} \equiv e^{- A( r_c \pi)} v_0
\end{equation}
Therefore, to generate the TeV scale from $\text{M}_{Pl}$, we only require $A(\pi rc)\sim 30$.This can be achieved for $b_0\sim \mathcal{O}(1-10)$, $\alpha\sim \mathcal{O}(0.1)$ and $r_c \sim \mathcal{O}(1-10)$.Again, we do not require large hierarchies between the fundamental parameters $\alpha,~r_c,~b_0,~k$.( where these parameters have been rendered dimensionless by suitable rescaling with Planck scale ). Therefore, in this model as well, it is possible to keep the scale of gravity unchanged while the scale of Higgs gets exponentially suppressed. 

\subsection{Modulus Stabilization}\label{stabilisation}
Goldberger and Wise developed a mechanism for stabilizing the modulus field, wherein they incorporated a scalar field in bulk, generating a potential for the modulus field, with a minimum at $r_c$. Also, from scalar-tensor theories of gravity, one can establish a mathematical equivalence between the scalar degrees of freedom and higher powers of $\text{R}$ in $\text{f(R)}$ theory. Here we show that the introduction of higher curvature  $\text{f(R)}$ gravity in the bulk  containing a $\text{R}^2$ term alone can stabilize the modulus field $T(x)$. 

From, \ref{(3)}, considering only the bulk action, the modulus potential is obtained by integrating out the extra dimension. 

\begin{equation}
\label{(29)}  
\rm V(T) =\rm  2 \int_{0}^{\pi }{d\phi\,Te^{-4\text{A($\phi\, ,T$)}}\text{B($\phi\, ,T$)}(\text{R}+\alpha \text{R}^2-\Lambda)}\Bigg|_{\rm potential ~part}
\end{equation}
Assuming the form of $\text{A($\phi\,  ,T$)}$ and $\text{B($\phi\,  ,T$)}$ from \ref{(17)} and \ref{(18)}, we can obtain the modulus potential from \ref{(29)}. Further, one can argue that the contribution coming from the alpha correction in the exponential part of the warp factor is negligible as compared to the contribution coming from the RS part. The modulus potential is finally obtained as follows:
\begin{equation}
\label{(30)}
\begin{split}
\rm V\left(T\right)\approx~&\rm \text{const}+ \text{P} ~\text{Ei}\left[-\frac{4 k \left(1+b_0 \pi  T \alpha \right)}{b_0 \alpha }\right]+\frac{e^{-4 k \pi  T}}{60466176}\Bigg\{C_1+C_2T+\frac{C_3}{\left(1+b_0 \pi  T \alpha \right)}\\
&-\rm \frac{C_4+C_5  T }{b_0{}^2 \left(1+b_0 \pi  T  \alpha \right){}^2}+\frac{C_6}{b_0+b_0{}^2 \pi T\alpha}+\frac{C_7+C_8T+C_9T^2}{\left(1+b_0 \pi  T \alpha \right){}^3}+\frac{C_{10}+C_{11}T+C_{12}T^2+C_{13}T^3}{\left(1+b_0\pi  T \alpha \right){}^4}\Bigg\}\\
\end{split}
\end{equation}
where $\text{Ei}(x)=-\int^{\infty}_{-x} dt e^{-t}/t$ and the details of the parameters have been provided in the appendix \ref{Appendix}.\\

    The following plots describe the functional dependence of the modulus potential on $\rm T(x)$. The modulus potential has a minimum for a stabilized value of $\rm T(x)$, which in turn, plays a crucial role in modulus stabilization.
\begin{figure}[hbt!]
\begin{subfigure}{.450\linewidth}
  \includegraphics[width=\linewidth]{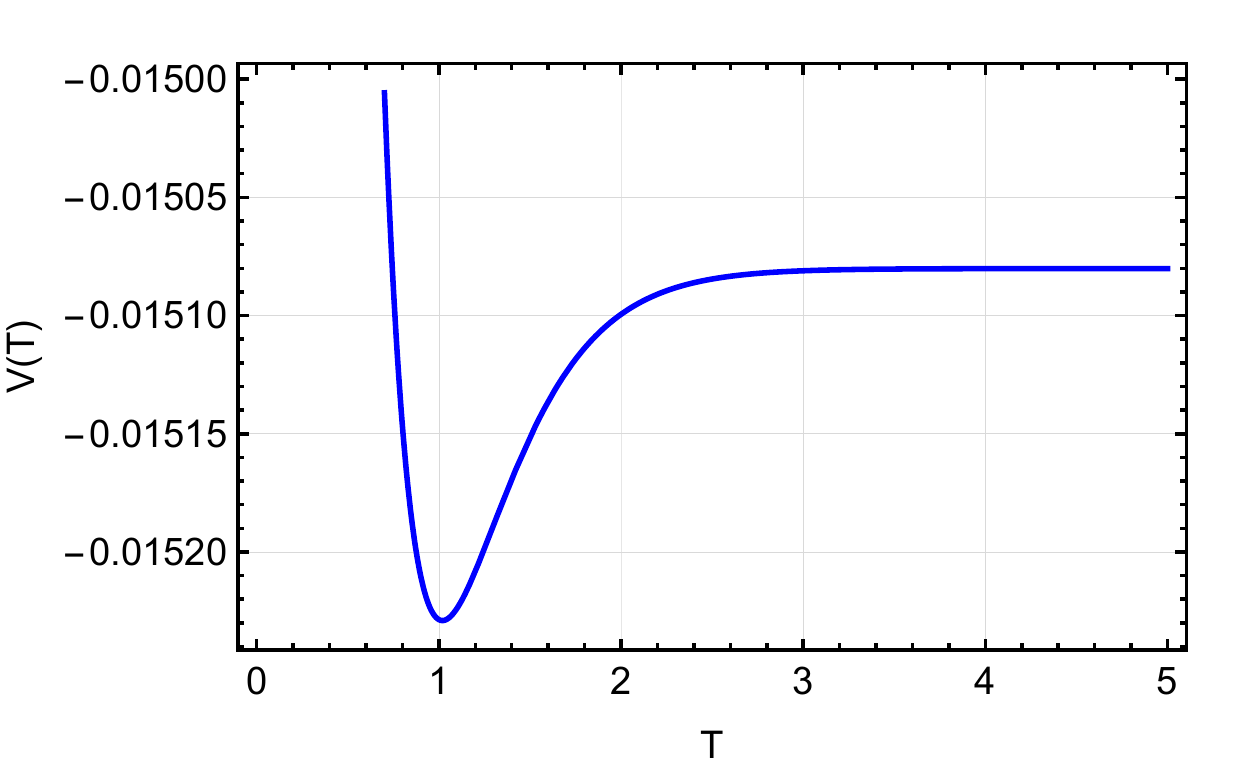}
  \caption{V(T) vs T for $\alpha=0.49$ and $b_{0}=1.19$}
  \label{fig2}
\end{subfigure}\hfill 
\begin{subfigure}{.440\linewidth}
  \includegraphics[width=\linewidth]{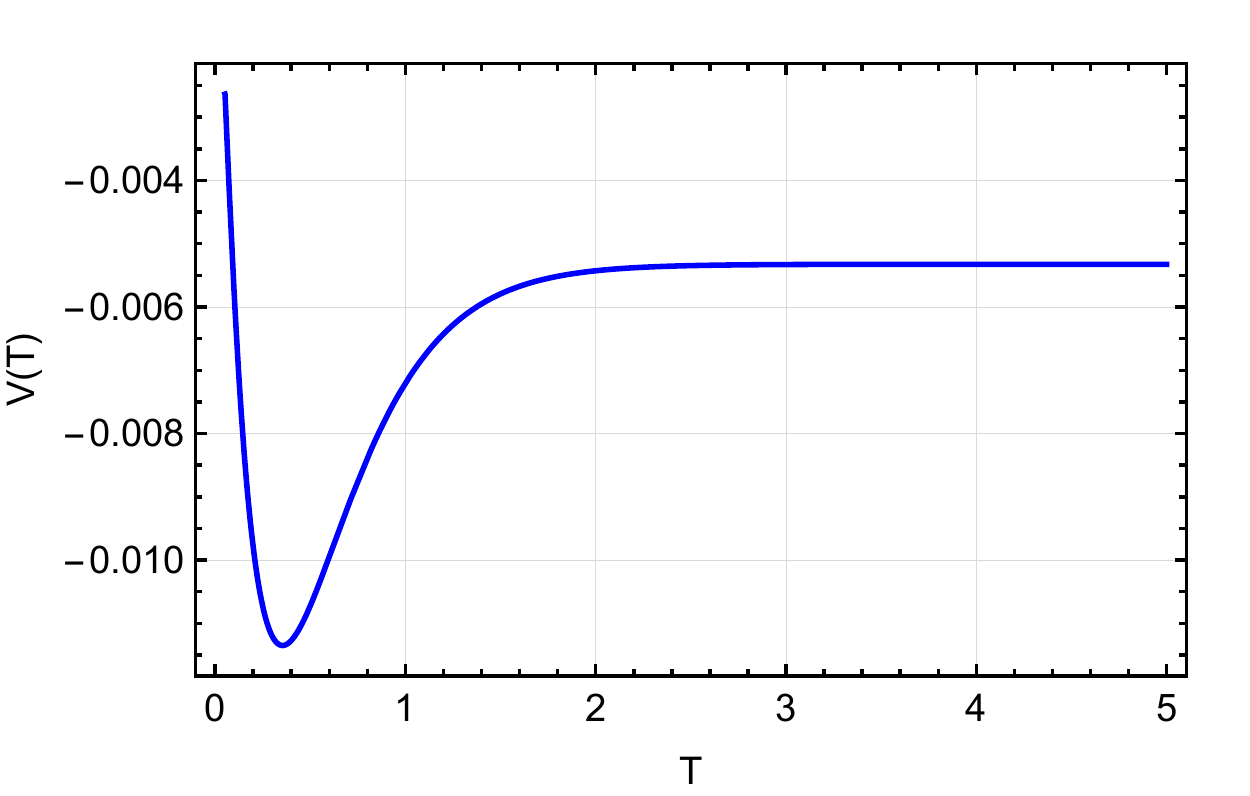}
  \caption{V(T) vs T for $\alpha=0.49$ and $b_{0}=4$}
  \label{fig3}
\end{subfigure}
\medskip 
\begin{subfigure}{.450\linewidth}
  \includegraphics[width=\linewidth]{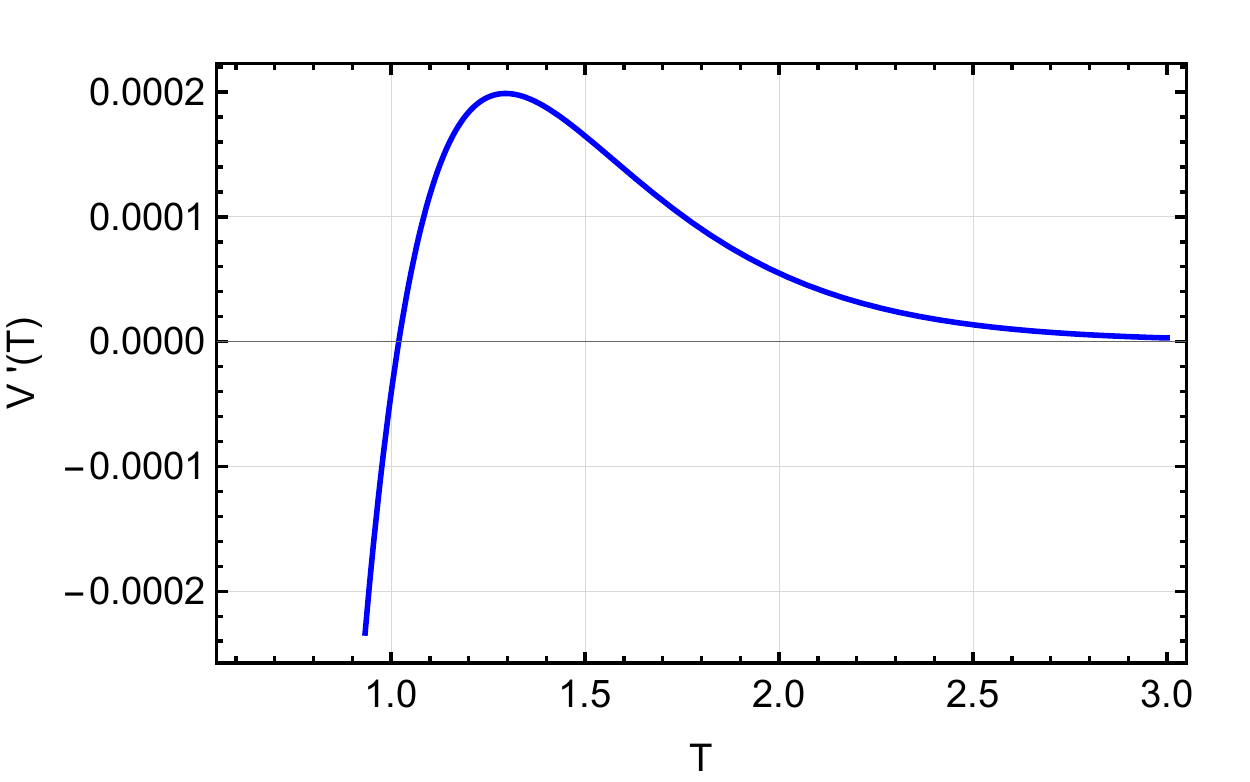}
  \caption{$V'(T)$  vs  T for $\alpha=0.49$ and $b_{0}=1.19$}
  \label{fig4}
\end{subfigure}\hfill 
\begin{subfigure}{.440\linewidth}
  \includegraphics[width=\linewidth]{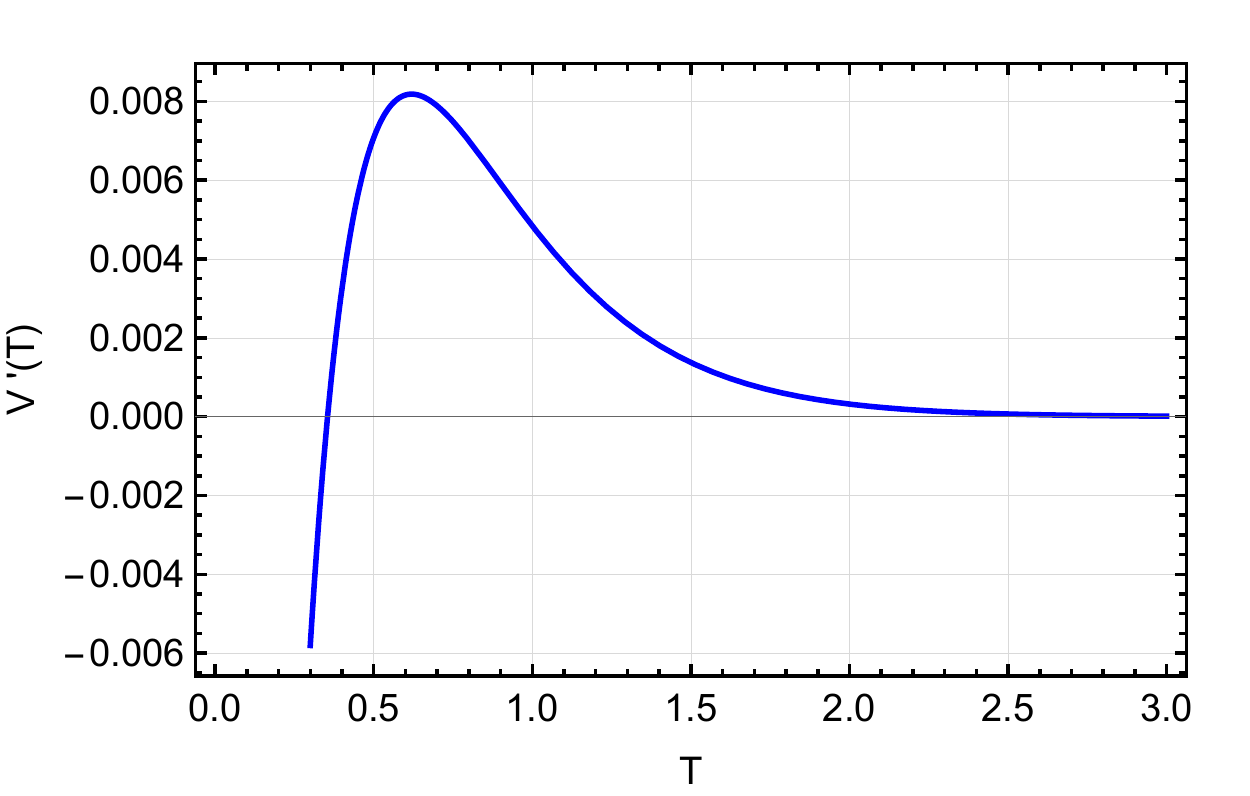}
  \caption{$V'(T)$ vs T for $\alpha=0.49$ and $b_{0}=4$}
  \label{fig5}
\end{subfigure}

\caption{Plots of modulus potential and its derivative for different values of $\alpha$ and $b_{0}$}
\label{fig:potential}
\end{figure}

\noindent

Our analysis reveals the following features:\\
\begin{itemize}
    \item Stabilisation can be achieved for $0<\alpha<1$, suggesting that our perturbative approach works.
   \item As $\alpha \xrightarrow{}0$ , stabilization no longer holds .
    \item To be consistent with $f'(R)>0$, we obtain the following bound on the model parameters i.e; $\alpha>0$ and $b_{0}>0$.
    \item Therefore, we have  $0<\alpha< 1$ and $b_{0}>0$ to obtain modulus stabilisation in ghost-free quadratic $\text{f(R)}$ gravity.
    \item For a fixed value of $b_{0}$, as the value of $\alpha$ increases, stabilization occurs for a smaller value of $T(x)$. 
    \item For a fixed value of $\alpha$, as the value of $b_{0}$ increases, the minima of the potential shifts, and now the stabilization occurs for a smaller value of $T(x)$. In \ref{fig2}, stabilization occurs at $\langle T(x) \rangle\approx 1$, whereas in \ref{fig3}, stabilization occurs at $\langle T(x) \rangle\ \approx 0.4$.
\end{itemize}
Furthermore, in \ref{fig4} and \ref{fig5} we have taken the derivative of the modulus potential, which further reinstates our claim about the occurrence of the minima. The stabilizing value of the radion field,  $\langle T(x) \rangle$ is obtained for $dV(T)/dT=0$.

In \ref{fig4} and \ref{fig5}, the critical points occur at $\langle T(x) \rangle\approx1$ and $\langle T(x) \rangle\approx0.4$, respectively, which in turn, correspond to the minima of the modulus potential as portrayed in \ref{fig2} and \ref{fig3}. On the other hand, on expanding the exponential up to the next order correction in $\alpha$ in \ref{(29)}, only the position of the minima gets shifted and now stabilization can be obtained for a smaller value of $\alpha$, for a given value of $b_{0}$. 

\begin{figure}[ht]           
             
             \ffigbox{\includegraphics[height=1.7in]{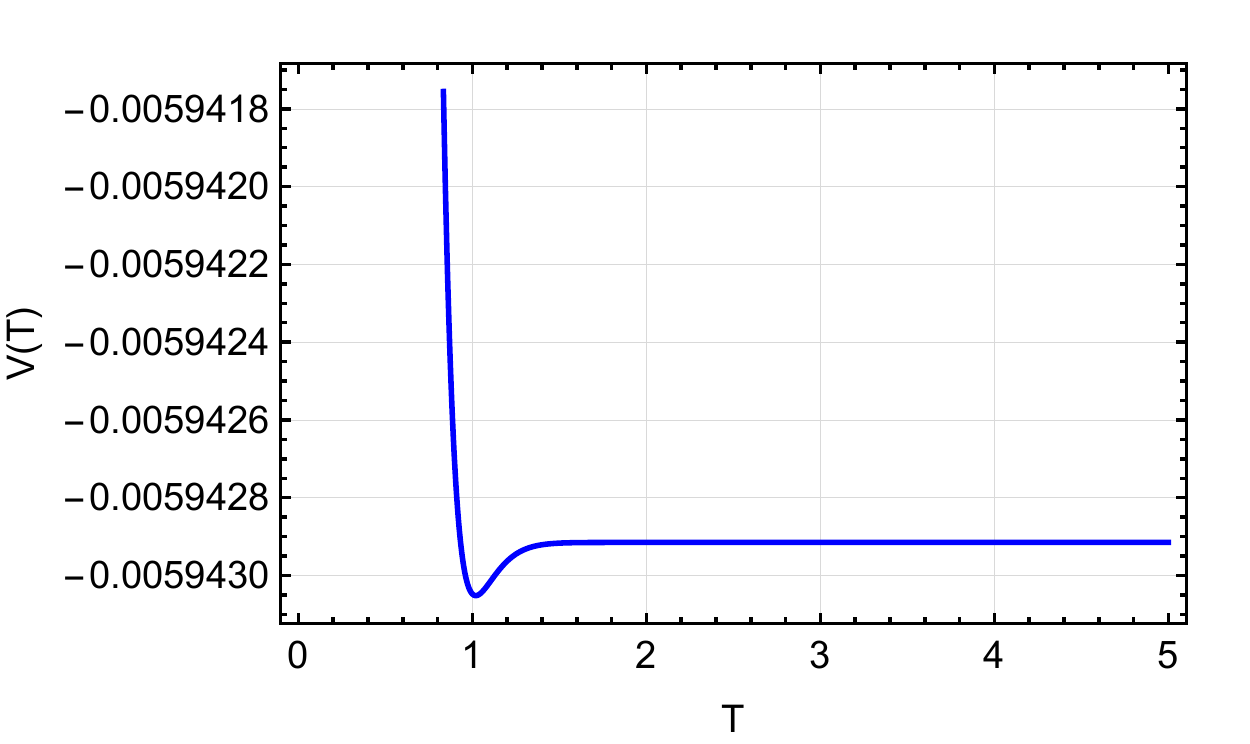}}{\caption{ Plot of V(T) vs T, obtained from numerical integration, for $\alpha=0.49$ and $b_{0}=1.19$} \label{fig7}}

        \end{figure}

Further, we can numerically integrate the modulus potential obtained in \ref{(29)}. Numerical integration yields the following plot. From \ref{fig7}, $\langle T(x) \rangle \approx 1$, same as that in \ref{fig2}, suggesting that our analytic estimate works. 
\FloatBarrier



\subsection{Swampland conjecture and Cosmological implications}\label{swampland}
\begin{figure}[hbt!]

  \includegraphics[height=1.7in]{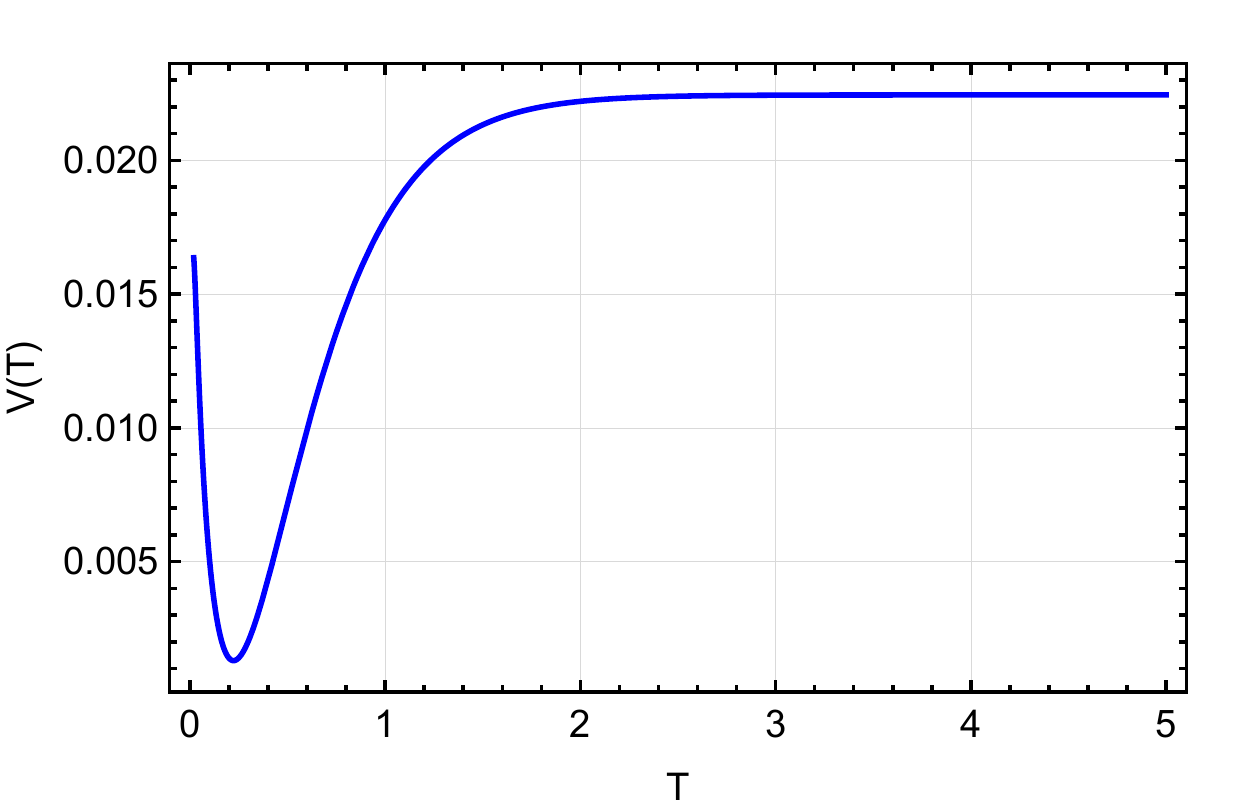}
  \caption{V(T) vs T for $\alpha=0.484$ and $b_{0}=22.35$}
  \label{swamp}
 
\end{figure}

Swampland constraints in string-inspired models  \cite{Ooguri:2006in}-\cite{Montefalcone:2020vlu} help to eliminate a large class of extra-dimensional theories. To explain the cosmological expansion 
in various epochs, the associated scalar field $T$ and its potential $\rm V(T)$ need to satisfy a set of constraints.  The constraints put conditions on $V(T)$ , the field range $\rm \Delta T$, and derivatives $\rm \partial_T V(T)$  and $\rm \partial_T ^2 V(T)$ . These constraints 
in turn are related to the equation of state. The swampland of string theory comprises consistent lower-dimensional effective field theories coupled to 
gravity. The slope conditions imply that the scalar field, such as $\rm T$ ( radion in this scenario ), must satisfy the conditions: 
$\rm V(T)>0$ and $\rm |\partial_T V(T)/V(T)|\sim $$\cal{O}$$(1)$. From \ref{fig3} and \ref{swamp}, it may be seen that for the appropriate choice of parameters one can satisfy the conditions on the scalar sector which emerges from the swampland
conjecture. The corresponding numerical estimate for our model is given below.
\begin{itemize}
    \item $\rm V(T)>0$ for $\alpha \approx 0.484$ and $21.4<b_0<23$ 
\end{itemize}
From \ref{swamp} it may be seen that the radion potential V(T) closely corresponds to the Starobinsky Inflaton potential  $\rm V(T) \sim \alpha ( 1- e^{-\sqrt{\alpha}T})^2$. As the radion scalar rolls down to the stable minimum, the scalar spectral index of the curvature perturbations $\rm n_s$ and the scalar to tensor ratio  r can now be determined from 
the corresponding slow-roll parameters for e-folding around $60$. It may further be noted that the radion scalar here rolls down to the vacuum with a non-vanishing
value for the ground state energy leading to a possible source of dark energy. A detailed analysis of these to determine the appropriate choices for the parameters of the theory will be considered in future work.

\FloatBarrier

\section{Conclusion}\label{Accretion_Conc}
Our results reported in this work reveal the following novel features.
Gravity action in space-time in general can contain curvature terms of various orders due to the underlying diffeomorphism invariance.
The terms containing n-th powers of curvature term are  suppressed by $\text{M}_{P}^{-2n} $ and therefore do not have much significance in our nearly flat universe. However in a higher dimensional space-time with a bulk cosmological constant of the order
of the Planck scale, such higher curvature terms are expected to have a significant contribution to determining the bulk geometry. 
This motivates us to generalize the earlier work on the warped geometry model  in a bulk action to include higher curvature $\text{f(R)}$ action.
We indeed obtain new warped geometry solutions with new  signatures in the effective theory on our 3-brane. Such signatures however
depend critically on the stable value of the modulus field which appears through various parameters in the effective lower dimensional theory. To extract an acceptable lower dimensional theory we, therefore, need to generate a modulus potential on the 3-brane.
It was shown earlier that only Einstein's action in the bulk fails to generate any such modulus potential and one needs to include an ad-hoc external stabilizing field to achieve the stabilization. Our work reported here brings out a remarkable and novel feature of a natural geometric stabilization mechanism of the modulus through the higher curvature gravity terms which is known to be equivalent to the presence of additional degrees of freedom in the bulk. The new warped geometry not only addresses the resolution of the gauge hierarchy problem successfully but also leads to possible new signatures on the brane such as graviton KK modes and their coupling to the standard model fields along with new physics of the radion field from the resulting modulus potential. 
Furthermore, this work opens up the study of various  cosmological implications discussed in \ref{swampland} along with a possible  mechanism of reheating at the end of inflation. We envisage addressing these in future work. This work thus brings out  a novel mechanism of the geometric radion stabilization mechanism which is at the root of all future studies of cosmological as well as collider-based observations in higher dimensional models.

\section{Acknowledgement}
Shafaq Elahi and Soumya Samrat would like to thank Dr.Tanmoy Paul and Dr.Sumanta Chakraborty for their valuable insights. SGE and SSM are supported by IACS MS- studentship.

\section{Appendix}
\subsection{5-D Ricci Scalar}\label{5D-Ricci-Definitions}
The expression for the 5-D scalar curvature is: \ref{scalar-curvature}
\begin{align}
\rm R= &\rm (1+\alpha  b_0T(x)|\phi|)e^{2T(x)|\phi| (\Tilde{k} +\beta T(x)|\phi|)}~{}^4\mathcal{R} -(\partial_\mu T(x))^2\frac{2|\phi|~ e^{2 ~T(x)~ |\phi|(\Tilde{k} +\beta  ~T(x)~ |\phi|)}}{T(x)~ (1+\alpha  b_0T(x)~ |\phi|)^3} \mathcal{K}(x,|\phi|) \\ \nonumber
&-\rm \partial_\mu\partial^\mu T(x)\frac{2 e^{2 T(x) |\phi| (\Tilde{k} +\beta  T(x) |\phi|)}}{T(x) (1+\alpha  b_0 T(x) |\phi|)}\mathcal{P}(x,|\phi|)
-\mathcal{Q}(x,|\phi|)
\end{align}
\begin{align}
 \mathcal{K}(x,|\phi|)=& 12 \beta ^2 ~T(x)^3 |\phi|^3 (1+\alpha  b_0~T(x)|\phi|)^3\\ \nonumber
 &+\rm 2 \Tilde{k}  (\alpha  b_0~T(x)|\phi|+1)^2 (2 ~T(x)~ |\phi|~ (3 \beta T(x)|\phi|~ (\alpha  b_0~T(x)~ |\phi|~+1)-\alpha  b_0)-1)\\ \nonumber
&\rm-2 \beta  ~T(x)~ |\phi|~ (7 \alpha  b_0~T(x)|\phi|+5)  (\alpha  b_0~T(x)~ |\phi|~+1)^2+3 \Tilde{k} ^2 ~T(x)~ |\phi|~ (\alpha  b_0~T(x)~ |\phi|~+1)^3\\ \nonumber
&\rm +2 \alpha  b_0(\alpha  b_0~T(x)~ |\phi|~ (2-\alpha  b_0~T(x)~ |\phi|~)+1)\\ \nonumber
\mathcal{P}(x,|\phi|)=&(T(x) |\phi| (\alpha  b_0 (3 T(x) |\phi| (\Tilde{k} +2 \beta  T(x) |\phi|)-2)+3 (\Tilde{k} +2 \beta  T(x) |\phi|))-1)\\ \nonumber
\mathcal{Q}(x,|\phi|)=&\frac{2}{~T(x)~ (\alpha  b_0~T(x)~ |\phi|~+1)^3}\left[ T(x)~ \left(20 \beta ^2 ~T(x)~^2 |\phi|~^2 \right.\right.\\ \nonumber
&\left.\left.\rm (\alpha  b_0~T(x)~ |\phi|~+1)+4 \beta  (5 \Tilde{k}  ~T(x)~ |\phi|~ (\alpha  b_0~T(x)~ |\phi|~+1)-1)+\Tilde{k}  (\alpha  b_0(5 \Tilde{k}  ~T(x)~ |\phi|~+2)+5 \Tilde{k} )\right)\right]
\end{align}
\subsection{Relationship between 4-D and 5-D Planck Scale}\label{4D5DAppendix}
In \ref{4Dgravity}, we obtained the following expression for $\cal{F}$:
\begin{equation*}
\rm    \mathcal{F}(r_c,\alpha,k,\text{M},b_0)=\rm e^{\Tilde{P}(r_c,\alpha,k,\text{M},b_0)}-\frac{1}{3 \sqrt{b_0}}10 \sqrt{\pi } \sqrt{\alpha } k^{5/2} e^{\Tilde{Q}(r_c,\alpha,k,\text{M},b_0)}\Big\{\textbf{erf}\left[\Tilde{N}(r_c,\alpha,k,\text{M},b_0)\right]-\textbf{erf}\left[\Tilde{S}(r_c,\alpha,k,\text{M},b_0)\right]\Big\}
\end{equation*}
where:
\begin{align*}
    \Tilde{P}=&-\rm \frac{1}{3} \pi  k r_c \left(3 \pi  b_0 \alpha r_c+\frac{20 \alpha  k^2}{\text{M}^2}+6\right)\\
    \Tilde{Q}=&\rm \frac{k \left(10 \alpha  k^2+3 \text{M}^2\right)^2}{9 b_0 \alpha  \text{M}^4}\\  
    \Tilde{N}=&\rm \frac{\sqrt{k} \left(10 \alpha  k^2+3 \text{M}^2\right)}{3 \sqrt{b_0} \sqrt{\alpha } \text{M}^2}\\
    \Tilde{S}=&\rm \frac{\sqrt{k} \left(3 \text{M}^2 (\pi  b_0 \alpha  r_c+1)+10 \alpha  k^2\right)}{3 \sqrt{b_0} \sqrt{\alpha } \text{M}^2}
\end{align*}
\subsection{Analytical Estimate for the potential}\label{Appendix}
The modulus potential was given by the following expression:
\begin{equation*} 
\rm V(T)\supset 2\int_{0}^{\pi }{d \phi~ T \, B(\phi\, ,T) e^{-4A(\phi\, ,T)}(\text{R}+\alpha \text{R}^2-\Lambda})
\end{equation*}
In \ref{stabilisation}, we obtained the following form of the modulus potential as given by \ref{(30)}:

\begin{equation*}
\begin{split}
V\left(T\right)\approx ~ &\text{const}+ \text{P} ~\text{Ei}\left[-\frac{4 k \left(1+b_0 \pi  T \alpha \right)}{b_0 \alpha }\right]+\frac{e^{-4 k \pi  T}}{60466176}\Bigg\{C_1+C_2T+\frac{C_3}{\left(1+b_0 \pi  T \alpha \right)}\\
&-\frac{C_4+C_5T}{b_0{}^2 \left(1+b_0 \pi T \alpha \right){}^2}+\frac{C_6}{b_0+b_0{}^2 \pi  T \alpha }+\frac{C_7+C_8T+C_9T^2}{\left(1+b_0 \pi  T \alpha \right){}^3}+\frac{C_{10}+C_{11}T+C_{12}T^2+C_{13}T^3}{\left(1+b_0\pi  T \alpha \right){}^4}\Bigg\}
\end{split}
\end{equation*}
In the above expression, the associated parameters are given by the following values:\\
\begin{align*}
 \text{const}=&-\frac{1}{60466176 k^2 b_0{}^3}\Biggl\{-b_0 \left(-16588800 k^5 \alpha +38400 k^4 \left(25 \sqrt{6}+108 b_0\right) \alpha ^2+209952 b_0{}^3 \alpha  (-12+25 \alpha )\right.\\
&+93312 k b_0{}^2 \left(-108+150 \alpha +125 \alpha ^2\right)-100 k^3 \alpha ^2 \left(625 \alpha +20736 b_0{}^2 \alpha +480 \sqrt{6} b_0 (216+5 \alpha )\right)\\
&+5 k^2 b_0\alpha  \left(311040 b_0{}^2\alpha ^3+625 b_0{}^2 (288+5 \alpha )\right)+192 \sqrt{6} b_0 \left(-11664+19440 \alpha +2700 \alpha ^2\right.\\
&\left.+125 \alpha ^3\right) 80 e^{\frac{4 k}{b_0 \alpha }} k^2 \left(829440 k^4-48000 \sqrt{6} k^3 \alpha +24300 b_0{}^2 \alpha  (-1+5 \alpha)\right.\\
&\left.+25 k^2 \alpha  \left(20736\sqrt{6} b_0+125 \alpha \right)-72 k b_0 \left(625 \alpha ^2+2592 \sqrt{6} b_0(-3+5 \alpha )\right)\right) \text{Ei}\left[-\frac{4 k}{b_0\alpha }\right]\Biggr\}
\end{align*}

\begin{doublespace}
\noindent\(\pmb{\text{}}\\
\pmb{}\\
{\text{P}= \frac{1}{7558272 b_0{}^3} \Biggl\{ 10 e^{\frac{4 k}{b_0 \alpha }} \left(829440 k^4-48000 \sqrt{6} k^3 \alpha +24300 b_0{}^2 \alpha  (-1+5 \alpha )+25 k^2 \alpha  \left(20736
\sqrt{6} b_0+125 \alpha \right)\right.}\\
{\left.-72 k b_0 \left(625 \alpha ^2+2592 \sqrt{6}b_0 (-3+5 \alpha )\right)\right)\Biggr\}}\\
{C_1=-\frac{1}{k^2}209952 b_0 \alpha  (-12+25 \alpha )+\frac{1}{k}10077696-13996800 \alpha -11664000 \alpha ^2}\\
{C_2=\frac{1}{k}10077696 b_0 \pi  \alpha -20995200 b_0 \pi  \alpha ^2}\\
{C_3=\frac{1}{b_0{}^2 \left(1+b_0 \pi  r_c \alpha \right)}16588800 k^3 \alpha }\\
{C_4=960000 \sqrt{6} k^2 \alpha ^2+4147200 k^2 b_0 \alpha ^2}\\
{C_5=960000 \sqrt{6} k^2 b_0 \pi  \alpha ^3}\\
{C_6=10368000 \sqrt{6} k \alpha ^2}\\
{C_7=\left(2073600 k +\frac{62500 k }{b_0 {}^2}+\frac{240000 \sqrt{6} k }{b_0 }\right)\alpha ^3}\\
{C_8=240000 \sqrt{6} k \pi  \alpha ^4+\frac{125000 k \pi  \alpha ^4}{b_0}}\\
{C_9=62500 k \pi ^2 \alpha ^5}\\
{C_{10}=11197440 \sqrt{6} \alpha -18662400 \sqrt{6} \alpha ^2-\left(2592000 \sqrt{6} +\frac{900000 }{b_0}\right)\alpha ^3-\left(120000 \sqrt{6}
+\frac{15625 }{b_0}+1555200 b_0 \right)\alpha ^4}\\
{C_{11}=33592320 \sqrt{6} b_0 \pi  \alpha ^2-55987200 \sqrt{6} b_0 \pi  \alpha ^3-\left(2700000 \pi  +5184000 \sqrt{6} b_0 \pi  \right)\alpha
^4-\left(31250 \pi  +120000 \sqrt{6} b_0 \pi  \right)\alpha ^5}\\
{C_{12}=33592320 \sqrt{6} b_0{}^2 \pi ^2 \alpha ^3-55987200 \sqrt{6} b_0{}^2 \pi ^2 \alpha ^4-\left(2700000 b_0 \pi ^2 +2592000 \sqrt{6} b_0{}^2
\pi ^2\right) \alpha ^5-15625b_0 \pi ^2 \alpha ^6}\\
{C_{13}=11197440 \sqrt{6} b_0{}^3 \pi ^3 \alpha ^4-18662400 \sqrt{6} b_0{}^3 \pi ^3 \alpha ^5-900000 b_0{}^2 \pi ^3 \alpha ^6}\)
\end{doublespace}
\end{document}